\documentstyle[twoside,fleqn,espcrc2,psfig]{article}
\newcommand{\beq}{\begin{equation}}
\newcommand{\eeq}{\end{equation}}
\newcommand{\bed}{\begin{displaymath}}
\newcommand{\eed}{\end{displaymath}}
\newcommand{\bea}{\begin{eqnarray}}
\newcommand{\eea}{\end{eqnarray}}

\renewcommand{\ni}{\noindent}

\renewcommand{\b}{\beta}

\newcommand{\m}{\mu}

\newcommand{\s}{\sigma}

\newcommand{\th}{\theta}

\newcommand{\oh}{\frac{1}{2}}

\newcommand{\dg}{\dagger}
\newcommand{\non}{\nonumber}

\newcommand{\rf}[1]{(\ref{#1})}
\newcommand{\ra}{\rightarrow}

\newcommand{\AmS}{{\protect\the\textfont2
  A\kern-.1667em\lower.5ex\hbox{M}\kern-.125emS}}

\title{Some Cautionary Remarks on Abelian Projection and Abelian Dominance}

\author{L. Del Debbio\address{Physics Dept., Univ. of Wales Swansea,
SA2 8PP Swansea, UK}, M. Faber\address{Inst. f{\"u}r Kernphysik, Tech. Univ.
Wien, 1040 Vienna, Austria}, J. Greensite\address{Physics Dept.,
San Francisco State Univ., San Francisco, CA 94132 USA}
\thanks{Talk presented by J. Greensite.  Work supported by the U.S. Dept. 
of Energy under Grant No. DE-FG03-92ER40711.},
\v{S}. Olejn{\'\i}k\address{Inst. of Phys., Slovak Acad. of Sci., 
842 28 Bratislava, Slovakia} }

\begin{document}

\begin{abstract}
   Some critical remarks are presented, concerning the abelian projection
theory of quark confinement.
\end{abstract}

\maketitle

\section{Introduction}

 Interest in the  abelian projection theory of quark confinement, proposed
by 't Hooft in 1981 \cite{tHooft}, has been increasing in recent years.
Very briefly, the idea is to 
select an ``abelian projection'' gauge which reduces the gauge symmetry of an
$SU(N)$ gauge theory to $U(1)^{N-1}$.  This choice enables one to identify
abelian gauge fields and magnetic monopoles.  The theory is that abelian
electric charge will then be confined by monopole condensation.
In this talk I would like to make some general remarks on these ideas,
and on the related concept of abelian dominance.

   Most of the recent efforts in the abelian projection theory have involved
lattice Monte Carlo simulations, and most of those
simulations have used a particular abelian projection gauge known
as the maximal abelian gauge, introduced by Kronfeld et al. \cite{Kronfeld}
in 1987.  On the lattice, for $SU(2)$ gauge theory, maximal abelian gauge
is obtained by maximizing the quantity
\begin{equation}
\sum_x \sum_\mu \mbox{Tr}[U_{\mu}(x) \sigma^3 U^\dagger_{\mu}(x)\sigma^3]
\end{equation}
This gauge choice has the effect of making link variables as diagonal as
possible, leaving a residual $U(1)$ symmetry.  Having made the maximal
abelian (or any other abelian projection) gauge choice, one can 
extract, from the full link variable $U_\m(x)$, a diagonal matrix $A_\m(x)$
which transforms as an abelian gauge field under the residual $U(1)$
symmetry:
\bea
           U &=& WA 
\non \\
           A &=& \left[ \begin{array}{cc}
                      e^{i\th} &  \\
                               & e^{-i\th} \end{array} \right]
\label{A}
\eea
(see also eq. \rf{proj} below).
Using the $A$ abelian gauge field, one may investigate (among other
things): (i) Creutz
ratios and Polyakov lines \cite{Suzuki}; (ii) monopole densities 
\cite{Kronfeld}; (iii) dual London relations \cite{Haymaker}; and (iv)
expectation values of monopole creation operators \cite{Pisa,ITEP}.
Of particular note is the work of the Kanazawa group \cite{Suzuki}, who
found that certain quantities such as the string tension, extracted
from Wilson loops constructed from the abelian link variables $A$ alone,
are close to the values obtained from the full link variables, a property
known as ``abelian dominance.''  A much more extensive list of recent work 
can be found in ref. \cite{Poli}.

   From these and other investigations, two conclusions have been drawn:

\begin{description}

\item{1.} The $U(1)$ gauge field looks like that of a dual superconductor.

\item{2.} The abelian projection theory of confinement is confirmed.

\end{description}
   
\ni The first conclusion is well supported by existing data.  My remarks
will be directed at the second conclusion, which has been inferred from
the first.

\section{Center Dominance}

   To motivate the first calculations I will discuss, let me recall
a theory of confinement which was popular in the late 1970's, namely
the $Z_N$ vortex condensation theory \cite{tHooft1,Mack}.  In this 
picture the vacuum state of an $SU(N)$ gauge theory is presumed to be
dominated by long vortices carrying multiples of $Z_N$ magnetic flux
(a closely related picture was developed somewhat earlier by the
Copenhagen group; c.f. ref. \cite{NBI}).  If the
area of a spacelike Wilson loop is pierced by $m$ such vortices, the
value of the loop is multiplied by a factor of $\exp(2im\pi/N)$.
The area law is attributed to random fluctuations in the number of
$Z_N$ vortices piercing the loop. 't Hooft introduced a singular
gauge transformation
operator, denoted $B(C)$, which creates a closed $Z_N$ vortex along curve
$C$.  A necessary condition for magnetic disorder is a perimeter law
behavior
\beq
<B(C)> \sim \exp[-\m L(C)]
\eeq
for the VEV of the vortex creation operator.  It was also suggested
that the confining dynamics of an $SU(N)$ gauge theory is described by
an effective $Z_N$ gauge theory \cite{Mack1}.

    The heyday of the vortex condensation theory preceded the widespread
use of lattice Monte Carlo methods, but in view of the numerical work
that has been done in recent years on the abelian projection theory, it 
is interesting to go back and conduct similar numerical experiments on 
the $Z_N$ vortex theory.  The general idea is that, just as maximal
abelian gauge brings the link variables $U$ of an $SU(2)$ gauge theory  
as close as possible to the $U(1)$ gauge fields $A$, so we would
like to go one step further and use the remnant $U(1)$ gauge freedom
to bring the $A$ gauge field as close as possible to a $Z_2$ gauge
field, with values $\pm I$.  The procedure is to begin by fixing to
maximal abelian gauge, with $A$ the diagonal matrix shown in eq. \rf{A}.
We then use the remnant $U(1)$ symmetry to maximize
\begin{equation}
      \sum_x \sum_\m \cos^2(\th_\m(x))
\end{equation}
leaving a remnant $Z_2$ symmetry.  This we call ``Maximal $Z_2$ Gauge.''
Then define, at each link,
\begin{equation}
       Z \equiv \mbox{sign}(\cos \th) = \pm 1
\end{equation}
which transforms like a $Z_2$ gauge field under the remnant symmetry.
``Center Projection'' $U \ra Z$, analogous to ``abelian projection''
$U \ra A$, replaces full link variables by the center element $ZI$,
in the computation of observables such as Wilson loops and Polyakov
lines.

   Results for Creutz ratios in the center projection, for maximal
$Z_2$ gauge, are shown in Fig. 1.  The data was taken for $SU(2)$ lattice
gauge theory in $D=4$ dimensions.  Lattice sizes were $10^4$ for
$\b \le 2.3$, $12^4$ at $\b=2.4$, and $16^4$ at $\b=2.5$.  The straight
line is the standard scaling function for the asympototic string tension
\beq
     \s a^2 = {\s \over \Lambda^2}({6\over 11}\pi^2 \b)^{102/121}
                 \exp[-{6\over 11}\pi^2 \b]
\eeq
with the value $\sqrt{\s}/\Lambda = 67$.

\begin{figure}
\centerline{\hbox{\psfig{figure=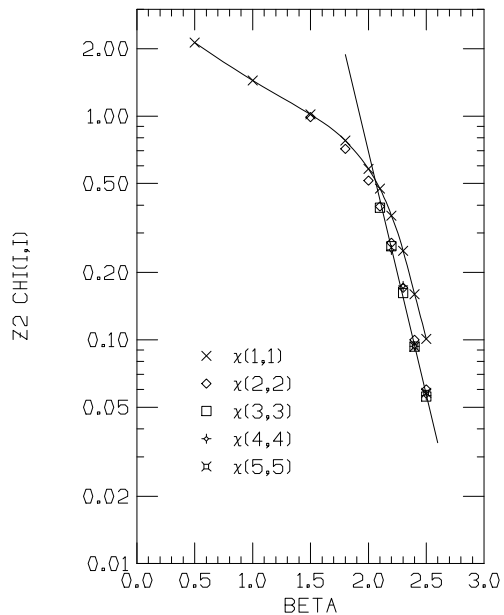,width=6.5cm,angle=90}}}
\caption[chi]{Creutz ratios from center-projected lattice
configurations.}
\label{chi}
\end{figure}

   What is remarkable about Fig. 1 is the fact that, from $\chi(2,2)$
onwards, the data for $\chi(R,R)$ at fixed $\b$ practically fall on top
of one another.  This is quite different from the standard curves, either
in the full theory or the usual abelian projection, where only the
envelope of the $\chi(R,R)$ data fits the scaling curve.  What it means
is that the center projection sweeps away the short-distance, $1/r$-type
potential, and the remaining linear potential is revealed almost from
the beginning.  This is seen quite clearly in Fig. 2, which displays
the data for $\chi(R,R)$ at $\b=2.4$ for the full theory (crosses),
the center projection (diamonds), and also for the $U(1)/Z_2$-projection
(squares).  The latter projection consists of the replacement $U \ra A/Z$ 
for the link variables.  We note that the center-projected data is virtually
flat, from $R=2$ to $R=5$, which means that the potential is linear in
this region, and appears to be the asymptote of the full theory.  It should
also be noted that abelian link variables with the center factored out, 
i.e. $U \ra A/Z$, appear to carry no string tension at all.

\begin{figure}
\centerline{\hbox{\psfig{figure=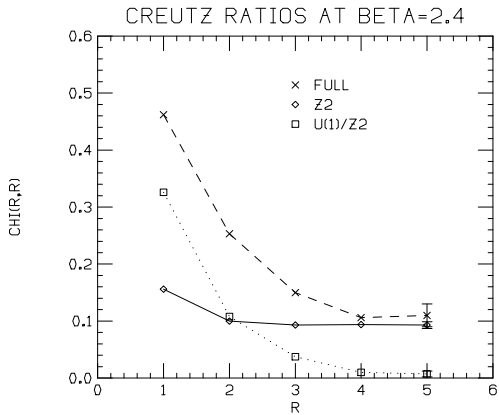,width=6.5cm,angle=90}}}
\caption[force]{Creutz ratios $\chi(R,R)$ vs. $R$ at $\b=2.4$, for
full, center-projected, and $U(1)/Z_2$-projected lattice 
configurations.}
\label{force}
\end{figure}

   We have not yet done a thorough investigation of finite temperature
effects in center projection.  Fig. 3, however, shows the Polyakov line
on a $6^3 \times 2$ lattice in center projection.  The phase transition,
indicated by a sudden jump in the data, appears to set in where it is
supposed to.

\begin{figure}
\centerline{\hbox{\psfig{figure=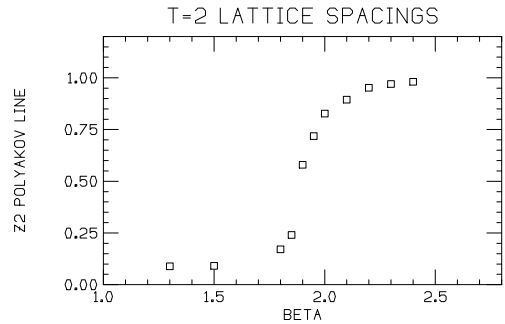,width=6.5cm,angle=90}}}
\caption[pol]{Polyakov lines vs. $\b$ in center projection, for
a $6^3 \times 2$ lattice.}
\label{pol}
\end{figure}

  $Z_2$ vortices are the {\it only} field configurations in $Z_2$
gauge theory, and we have found that the $Z_2$ configurations extracted
from $SU(2)$ lattice gauge theory give the full asymptotic string
tension (``center dominance'').  This data could certainly be 
taken as evidence 
in favor of the vortex condensation theory; in fact, various arguments 
in favor of the abelian projection theory can be reused to support the vortex 
condensation theory.  This recycling of arguments requires only a slight change
in terminology: replace ``maximal abelian gauge,'' ``abelian projection,''
and ``abelian dominance,'' by ``maximal $Z_2$ gauge,'' ``center projection,''
and ``center dominance,'' respectively, with the relevant topological
configurations being vortices rather than monopoles.  One would focus
on the VEV of vortex creation operators 
$<B(C)> \sim e^{-\m L(C)}$ rather than monopole creation operators
$<\Phi_M> \ne 0$, and the QCD ground state would resemble a 
``spaghetti vacuum'' of vortices, rather than a dual superconductor.

   Now if abelian dominance suggests that monopole condensation is
the confinement mechanism, while center dominance points instead to vortex
condensation, which mechanism should one believe in?  It could be argued
that while abelian dominance shows the abelian link to be the
crucial component of the full link, center dominance shows the
center variable to be the crucial variable of the abelian link, and therefore
center dominance is somehow the more basic phenomenon.  We would
not make this argument, however.  The reason is that neither the
vortex condensation theory, nor the abelian projection theory, gives an
adequate explanation of the potential between static sources in higher
group representations.

\section{Casimir Scaling}

   It has been found in many lattice Monte Carlo investigations, in
both $D=3$ and $D=4$ dimensions and for both $SU(2)$ and $SU(3)$ gauge 
groups, that the string tension of static sources in representation
$R$ of the gauge group is approximately proportional to the quadratic
Casimir $C_R$ of the representation \cite{AOP}.  This result holds with an
accuracy of about $10$\%, from the onset of confinement to the onset
of color screening.  In particular, for the $SU(2)$ gauge group,
\beq
  \s_j \approx {4\over 3}j(j+1) \s_{1/2} ~~~~~\mbox{intermediate region}  
\eeq
while asymptotically (after color-screening)
\beq
    \s_j \ra \left\{ \begin{array}{cl}
                     \s_{1/2} & j=\mbox{~half-integer} \\
                        0     & j=\mbox{~integer} \end{array} \right.
\eeq
This ``Casimir scaling'' of the string tension at intermediate distances
is easily derived at strong-couplings from either the
Kogut-Susskind Hamiltonian or the heat-kernel action.  It can also be
derived in $D=2$ dimensions, at weak couplings, for any lattice action.
It is not obvious why Casimir scaling persists at weaker couplings
in 3 and 4 dimensions, but probably the explanation is connected to
the concept of dimensional reduction, introduced in refs.
\cite{Me,Poul}, which would allow us to infer Casimir scaling in 3 and
4 dimensions from the 2-dimensional result.
In any case,  approximate Casimir scaling of string-tensions
up to the onset of color-screening is a numerical fact.  The question 
is whether this fact is consistent with either the vortex condensation
or the abelian projection theory.

   Consider first the vortex theory, and quarks in the adjoint representation.
The problem is that adjoint quarks are neutral (i.e. invariant) with respect
to $Z_2$ gauge transformations; as a consequence, adjoint Wilson loops
are unaffected by insertion or removal of a $Z_2$ vortex.  Thus, fluctuations
in the number of vortices piercing the loop cannot possibly be
responsible for an area law for adjoint loops; the vortex theory is
not at all compatible with Casimir scaling at intermediate distances.

   There is a similar problem for the abelian projection theory.
According to this theory it is the abelian charge, singled out by the
unbroken Cartan subalgebra, which is confined.  The adjoint representation
is $j=1$; the $m=+1$ and $m=-1$ color components have double abelian
charge ($++$ and $--$, respectively) as compared to the $+/-$ abelian
charge of the fundamental representation.  The $m=0$ component, however
is uncharged w.r.t the $U(1)$ subgroup, {\it and this neutrality holds
regardless of color screening}.  This means that there is no
apparent mechanism for string formation between $m=0$ quark components,
and therefore no reason for an area law; $\s_{j=1}=0$ would be expected
at all scales.  Similarly, for the $j=3/2$ representation, the
$m=\pm {3\over 2}$ components have triple abelian charge, while the 
$m=\pm \oh$ components have single abelian charge, the same as the
fundamental representation.  One therefore expects that $\s_{3/2}=\s_{1/2}$
prior to color screening, from the onset of confinement.

   It is really a qualitative point that is being made here.  If the
confining force is only sensitive to abelian charge, then the quark 
components with the lowest abelian charge should dominate the Wilson loops.
This point is illustrated by making the same abelian projection $U\ra A$
for the higher representation loops that has been used for the
fundamental loops.  In this illustration, one finds \cite{Smit}
\bea
     <W_j(C)> &=& <\mbox{Tr}\exp[i\oint dx^\m A_\m^a T^j_a]>
\non \\
              &\ra& <\mbox{Tr}\exp[i\oint dx^\m A_\m^3 T^j_3]>
\non \\
              &=&  \sum_{m=-j}^j <\exp[im\oint dx^\m A_\m^3]>
\non
\eea
so that in particular
\bed
     <W^{ab.}_1(C)> = 1 + <e^{i\oint A^3}> + <e^{-i\oint A^3}>
\eed
which implies $\s_{j=1}=0$ for abelian-projected configurations; and
\bea
  <W^{ab.}_{3/2}(C)> &=& <e^{i\oh\oint A^3}> + <e^{-i\oh\oint A^3}>
\non \\
   &+& <e^{i{3\over 2}\oint A^3}> + <e^{-i{3\over 2}\oint A^3}>
\non
\eea
implying $\s_{j=3/2}=\s_{j=1/2}$.  Again, these results have nothing to do
with color-screening.

   We conclude that confinement of {\it abelian} charge in an 
$SU(2)$ gauge theory would imply
\beq
    \s_{j=1}=0 ~~~~\mbox{and}~~~~ \s_{j=3/2} = \s_{j=1/2} 
\label{pred}
\eeq
from the confinement scale onwards, which disagrees with existing numerical
data.  However, the argument presented is 
rather qualitative, and it is obviously desirable to check eq. \rf{pred}
in a confining non-abelian gauge theory, where we are confident
that confinement is indeed due to abelian monopole configurations.

\section{The $3D$ Georgi-Glashow Model}

   The Georgi-Glashow model is an $SU(2)$ gauge theory with a Higgs
field in the adjoint representation.  It has been argued pursuasively
by Polyakov \cite{Polyakov} that in D=3 dimensions, 
confinement in the Higgs phase is
due to 't Hooft-Polyakov monopoles, which of course are instantons
in three dimensions.  We are therefore able to check that confinement
of abelian charge leads to eq. \rf{pred} by lattice Monte Carlo simulation.

   The lattice action for the Georgi-Glashow model is \cite{Nadkarni}
\bea
     S &=& \oh \b_G \sum_{plaq} \mbox{Tr}[UUU^\dg U^\dg]
\non \\
       &+& \oh \b_H \sum_{n,\m} \mbox{Tr}[U_\m(n)\phi^\dg(n+\m) U^{\dg}_\m(n)
                                                 \phi(n)]
\non \\
       &-& \sum_n \left\{ \oh \mbox{Tr}[\phi \phi^\dg]
           + \b_R \left(\oh \mbox{Tr}[\phi \phi^\dg]-1 \right)^2 \right\}
\non
\eea
Define observables
\bea
         R &=& <\mbox{Tr}[\phi \phi^\dg]>^{1/2}
\non \\
         Q &=& \oh <\mbox{Tr}[U_\m(n) \s^3 U^\dg_\m(n) \s^3]>
\non 
\eea
in unitary gauge $\phi=\rho \s^3$.
A jump in these two quantities is an indication of a transition
from the symmetric phase to the Higgs phase.  Our investigation has only
been at fixed values $\b_G=2,~\b_R=0.01$, and we look for the Higgs transition
by varying $\b_H$.  Figure 4 locates the Higgs transition near
$\b_H=0.45$ from a jump in $Q$.  A similar jump in the value of the
$R$ observable is found at this point.

\begin{figure}
\centerline{\hbox{\psfig{figure=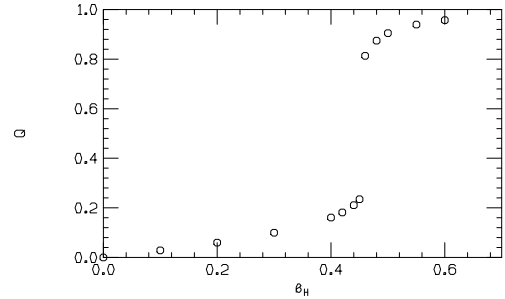,width=6.5cm,angle=90}}}
\caption[q]{Q vs. $\b_H$ in the D=3 Georgi-Glashow model.}
\label{q}
\end{figure}


   Figure 5 shows the behavior of the $\chi(2,2)$ Creutz ratios
for fundamental and adjoint loops
in the same range of couplings.  Note that prior to the transition
the adjoint value is roughly double the fundamental, while after
the transition the adjoint value is almost vanishing.

\begin{figure}
\centerline{\hbox{\psfig{figure=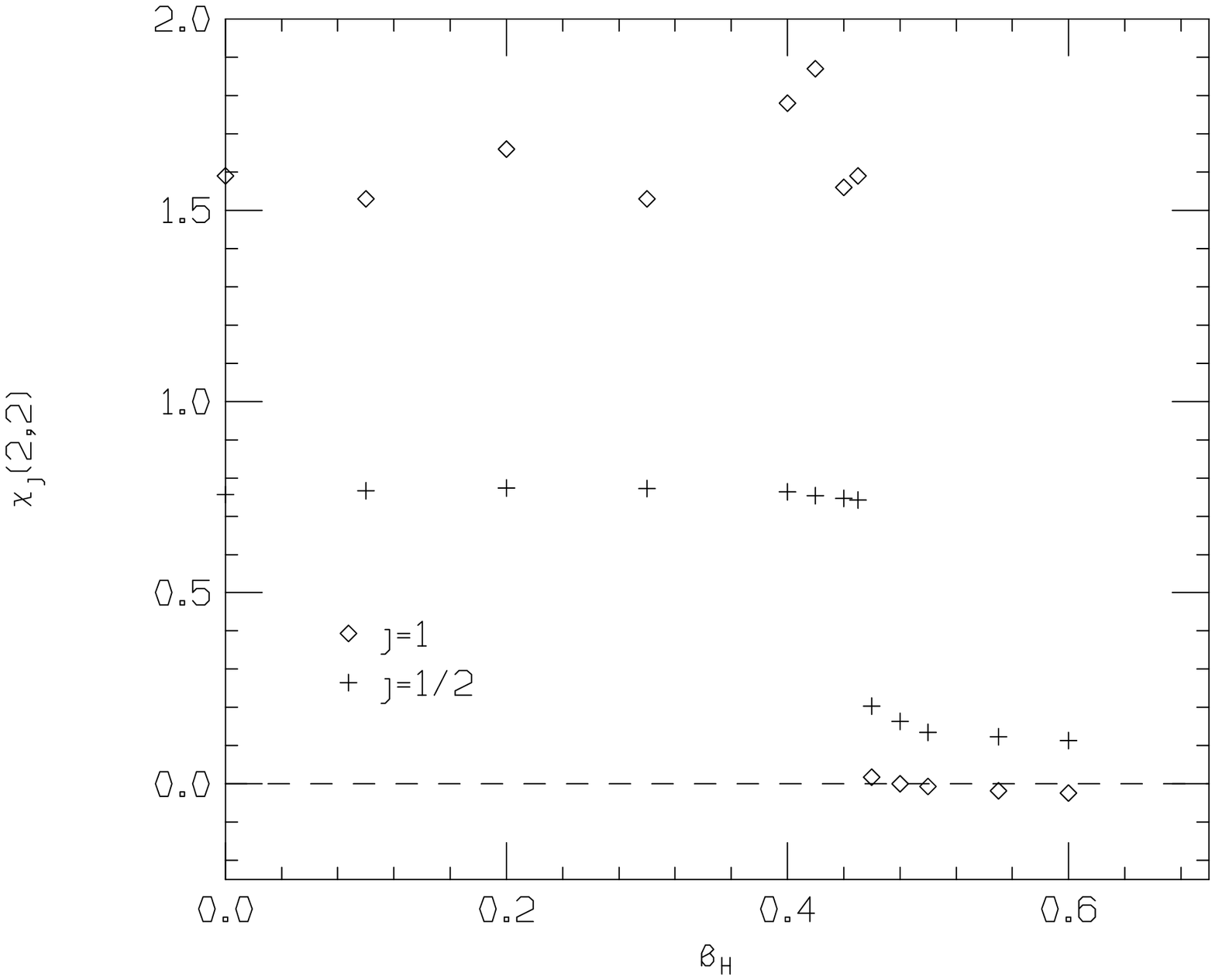,width=6.5cm,angle=90}}}
\caption[c22]{$\chi(2,2)$ Creutz ratios for fundamental and
adjoint loops, in the D=3 Georgi-Glashow model.}
\label{c22}
\end{figure}
   
   Creutz ratios in the Higgs phase at $\b_H=0.46$, which is just past 
the transition, are shown in Fig. 6.  The string tension for the
adjoint loop is clearly consistent with zero, with Creutz ratios
actually going negative at $I=3$. The data is also consistent
with $\s_{1/2} \approx \s_{3/2}$.  So this example does seem to support
the reasoning leading to eq. \rf{pred}.

\begin{figure}
\centerline{\hbox{\psfig{figure=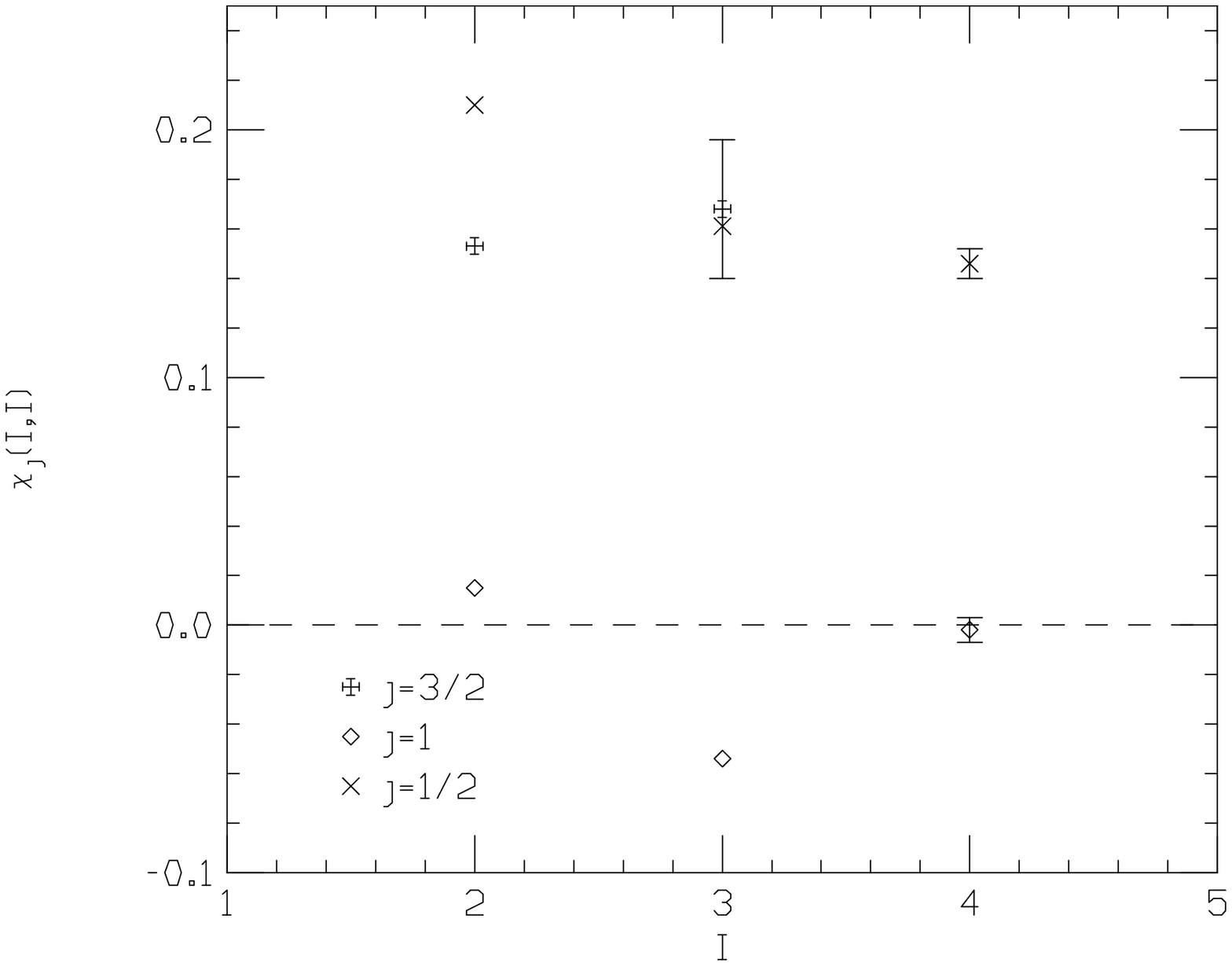,width=6.5cm,angle=90}}}
\caption[cjj]{$\chi_j(I,I)$ vs. $I$.  Creutz ratios for $j=\oh,1,{3\over 2}$,
just inside the Higgs phase of the D=3 Georgi-Glashow model.}
\label{cjj}
\end{figure}

   For comparison, let us return to pure Yang-Mills theory in $D=3$
dimensions.  It is found that the $\chi_j(I,I)$ ratios, extracted from
loops computed in abelian-projected configurations, are similar
to those shown in Fig. 6.  In contrast, Casimir scaling is quite evident
for ratios obtained from the full, unprojected
link configurations.  Space limitations prevent displaying the relevant
data here, but it may be found in Figures 1 and 2 of ref. \cite{Us}. 

   To summarize:
\begin{description}
\item[1.] In pure SU(2) gauge theory,
\bed
       \s_{j=1} \approx {8\over 3} \s_{j=1/2} ~~~~~~~~~
       \s_{j=3/2} \approx 5 \s_{j=1/2}
\eed
whereas,
\item[2.] In a theory where monopoles are known to drive the confinement
mechanism (D=3 Georgi-Glashow, Higgs phase)
\bed
       \s_{j=1} \approx 0  ~~~~~~~~~~~~
       \s_{j=3/2} \approx  \s_{j=1/2}
\eed  
\item[3.] In pure SU(2), with abelian projection in maximal abelian gauge,
\bed
       \s_{j=1} \approx 0 ~~~~~~~~~~~~
       \s_{j=3/2} \approx  \s_{j=1/2}
\eed
The result $\s_{j=1} \approx 0$ is simply due to the neutrality of
the $m=0$ adjoint quark component.
\end{description}
In connection with abelian-projected configurations, it should be noted 
that the contributions $<\exp[\pm i\oint A^3]>$ to the abelian adjoint loop,
\bed
     <W^{ab.}_1(C)> = 1 + <e^{i\oint A^3}> + <e^{-i\oint A^3}>
\eed
coming from the double-charged components $m=\pm 1$, {\it do} have an area
law, and in fact the string tension of these components is quite
close to that of the full, unprojected loops.  This could be taken as
evidence of some form of abelian dominance \cite{Poulis}, but that is somewhat
beside the point we are making here.  Confinement of the double-charged
adjoint quark components, in the abelian projection picture, is not in doubt.
The real issue is how the
linear potential can act on the abelian neutral ($m=0$) component, if 
only abelian charge is sensitive to the confining force. 

\section{XY-Maximal Abelian Gauge}

   Abelian dominance, at least for certain quantities, can be impressive. The 
general rule is to fix to maximal abelian gauge, do the $U\ra A$ abelian
projection 
\beq
      U = a_0 I + i\vec{a} \cdot \vec{\s} ~~~ \longrightarrow ~~~
      A = {a_0 I + i a_3 \s^3 \over \sqrt{a_0^2 + a_3^2} }
\label{proj}
\eeq
and then calculate observables with projected $A$ configurations.

    A skeptic, however, might argue that the maximal abelian gauge
choice forces most of the quantum fluctuations into the diagonal part of the
link.  Then perhaps it is not really surprising that the diagonal component
alone can reproduce various observables with reasonable accuracy;
the underlying reason may be more a matter of kinematics than dynamics.

  To investigate this issue, we ask the question: What if one chooses 
a gauge such that only links the XY-plane are as diagonal as possible, i.e.
\beq
      \sum_x \sum_{\m=1}^2 \mbox{Tr}[\s_3 U_\m \s_3 U^\dagger_\m]
           ~~~~~~~ \mbox{is maximized}
\eeq
and then abelian project? Do we find abelian dominance? And what if we 
``abelian project,'' via eq. \rf{proj}, without fixing any gauge at all?  
The answers are as follows:
\begin{figure}
\centerline{\hbox{\psfig{figure=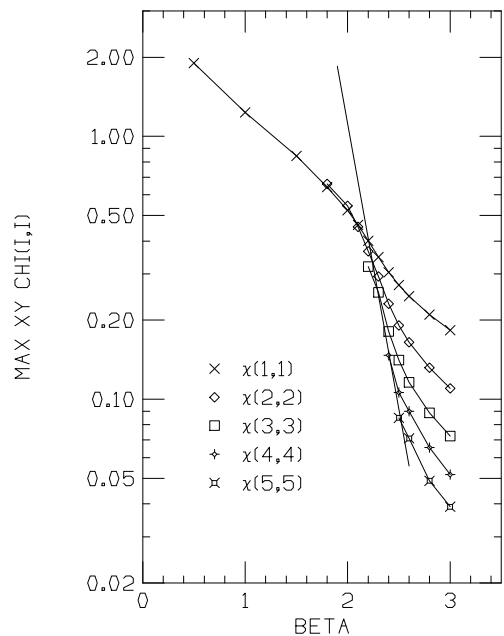,width=6.5cm,angle=90}}}
\caption[mxy]{Creutz ratios vs. $\b$, extracted from abelian loops in
the XY-plane in XY-maximal abelian gauge} \nopagebreak 
\label{mxy}
\end{figure} 
\nopagebreak
\begin{description}
\item{1.} Abelian-projected loops in the
XY-plane exhibit abelian dominance. 
\end{description}
Figure \ref{mxy} shows a plot of Creutz ratios vs. $\beta$, extracted from
abelian-projected loops in the XY-plane.  It has the standard form;
the envelope of $\chi(I,I)$ appears to fit a scaling curve, which is
shown with the value $\sqrt{\s}/\Lambda = 85$.  

\begin{description} 
\item{2.} There is no obvious abelian dominance
in the ZT-plane.  In fact, abelian loops in the ZT-plane are indistinguishable
from ``abelian-projected'' loops {\it with no gauge-fixing whatever!}
\end{description}
\begin{figure}
\centerline{\hbox{\psfig{figure=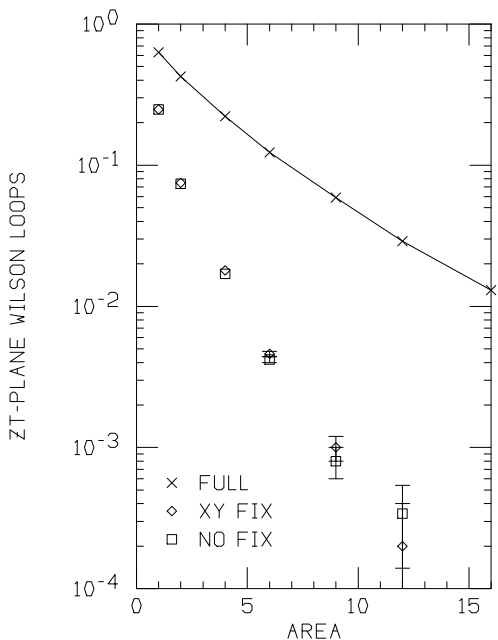,width=6.5cm,angle=90}}}
\caption[mzt]{Loop values vs. Area at $\b=2.4$, extracted from loops in
the ZT-plane.  Crosses show the full,
gauge-invariant values; diamonds are values for projected configurations
in XY-maximal abelian gauge; squares are values for projected configurations
with no gauge-fixing at all.} \nopagebreak
\label{mzt}
\end{figure}
\nopagebreak
Figure \ref{mzt} shows loop values vs. area at $\b=2.4$, for abelian projected 
loops in the ZT-plane (diamonds).  For comparison, loop values for the
full, unprojected configurations are also shown (crosses).  It is clear
that the loop values for projected configurations drop like a stone,
as compared to the full values (or compared to the projected loop 
values in the XY-plane).
Because of the rather large error bars, Creutz ratios taken from this 
data are not very meaningul.  What is of more significance is to compare
the loop values taken from projected configurations in the XY-maximal
abelian gauge, with configurations also projected according to eq. \rf{proj} 
but obtained using no gauge fixing whatever (squares).  It is clear 
that in the ZT-plane, in contrast to the XY-plane, the 
loop values of projected 
configurations are completely insensitive to the presence or absence of 
the XY-maximal abelian gauge fixing.

   This example supports the skeptical view.  Since we see abelian dominance
in the plane where links are nearly diagonal, and don't see it in
other directions, it suggests that abelian dominance is simply a consequence 
of having nearly diagonal links, and not necessarily evidence in favor of the
abelian projection theory.

\section{Conclusions}
  
   To summarize: we have found ``center dominance'' in maximal $Z_2$
gauge.  To the extent that abelian dominance supports the abelian projection
theory of confinement, center dominance supports the vortex condensation
theory. 

  Neither theory seems to explain, even qualitatively, the existence 
of a linear potential between adjoint quarks up to color-screening, let 
alone the approximate Casimir scaling of string tensions.  

  We have also seen that it is possible to choose a gauge 
(the XY-maximal abelian gauge), in which one sees ``abelian dominance'' in 
the XY-plane, but not in the ZT-plane.  What this suggests is that abelian 
dominance in maximal abelian gauge could be an artifact of setting links 
almost diagonal, rather than a definite indication
of the underlying confinement mechanism.

\end{document}